\documentclass[ps,prd,preprint,
superscriptaddress,preprintnumbers,eqsecnum,
noshowpacs,nofootinbib,nobibnotes]{revtex4}
\usepackage{amsfonts,bm}
\usepackage{graphicx}
\usepackage{color}

\newcommand{\be}{\begin{equation}}
\newcommand{\ee}{\end{equation}}

\def\operhs{\Upsilon}
\def\homo{\kappa}
\def\Y{\Xi}
\def\fg{\mathrm{I}\!\Gamma}

\def\diff{\mathrm{d}}

\begin{document}

\

\vskip 1.5 truecm

\title{Non-trivial Backgrounds\\ in (non-perturbative) Yang-Mills Theory\\ by the Slavnov-Taylor Identity}

\author{A Quadri}
\email{andrea.quadri@mi.infn.it}

\affiliation{Universit\`a degli Studi di Milano and INFN,
Sezione di Milano,\\ via Celoria 16, I-20133 Milan, Italy}

\vskip 0.3 truecm
\begin{abstract}
\noindent
 We show that in the background field method (BFM)
quantization of Yang-Mills theory the dependence of the
vertex functional on the background field is controlled
by a canonical transformation w.r.t. the Batalin-Vilkovisky
bracket, naturally associated with the BRST symmetry of the 
theory. 
Since it only relies on the Slavnov-Taylor  identity
of the model, this result
holds both in perturbation theory and in the
non-perturbative regime. 
It provides a general consistent framework for the
systematic implementation of the BFM
in non-perturbative approaches to QCD, like e.g.
those based on the Schwinger-Dyson equations
or the lattice, in the presence
of topologically non-trivial background configurations.
The analysis is carried out in an arbitrary (background) $R_\xi$-gauge.

\vskip 4.5 truecm
\noindent Prepared for Quantum Theory and Symmetries 7 \\
August 7-13, 2011, Prague.
\end{abstract}


\maketitle

\section{Introduction}

\noindent
 The Background Field Method (BFM)~\cite{DeWitt:1967ub,Abbott:1980hw} has played an important role in studying the properties of non-Abelian gauge theories.
In perturbative computations within the BFM, one has the
main advantage that gauge invariance with respect to the background field 
at the quantum level, encoded in the so-called background Ward identity, can be exploited in order to obtain
linear relations between 1-PI amplitudes.
This is in contrast with
the more complicated relations among 1-PI Green functions arising from the Slavnov-Taylor identity, which, unlike the background Ward identity, is bilinear in the vertex functional. 

Since perturbatively the BFM and the usual perturbation theory based on the Gell-Mann and Low's formula give
the same results for physical gauge-invariant observables ~\cite{Becchi:1999ir,Ferrari:2000yp}, the BFM can be used to significantly simplify computations
in several applications, ranging from perturbative calculations in Yang-Mills theories~\cite{Abbott:1980hw,Ichinose:1981uw} and in the Standard Model~\cite{Denner:1994xt,hep-ph/0102005} to gravity and supergravity calculations~\cite{Gates:1983nr}.

Outside ordinary perturbation theory, the BFM has been applied as a prescription for calculating to any order the $n$-point Green functions of the pinch technique~\cite{Cornwall:1981zr, Cornwall:1989gv} in the approach based on the (non-perturbative) Schwinger-Dyson equations~\cite{Binosi:2002ft,Binosi:2007pi}. 
In this context one makes use of the background-quantum identities ~\cite{Grassi:1999tp,Binosi:2002ez} relating Green functions involving a given combination of quantum and background fields with the same  functions where one of the background fields has been replaced by its quantum counterpart.

In the two-point sector of (pure) $SU(N)$ Yang-Mills theory these identities are important in controlling the IR dynamics of the
gluon and ghost propagators. The Schwinger-Dyson equation for the background gluon propagator can be truncated
gauge invariantly by exploiting the block-wise transversality of its gluon and ghost one- and two-loop dressed contributions~\cite{Binosi:2007pi,Aguilar:2006gr}. The solution of this equation can be then related to the conventional one through the corresponding two-point background quantum identity; the result is  a gauge-artifact-free propagator that can be meaningfully compared to the  high quality {\it ab-initio} lattice gauge theory computations currently available~\cite{Aguilar:2008xm}.

The IR properties of the gluon and ghost propagators have been extensively studied
by the Schwinger-Dyson equations \cite{Binosi:2007pi,arXiv:0801.2721,arXiv:0810.1987}.
The comparison between these continuum studies and lattice data provides convincing evidence that in the Landau gauge the
dressing functions of the gluon and ghost propagators tend in the deep IR to a finite, non-vanishing value~\cite{Cucchieri:2007md,Bogolubsky:2007ud},
thus supporting the mechanism of a dynamically generated gluon mass ~\cite{Cornwall:1981zr}.

This is 
in contrast with the predictions of the IR divergent ghost dressing function of the Kugo-Ojima confinement scenario~\cite{Kugo:1979gm}
and the IR divergent ghost dressing function and the IR vanishing gluon propagator typical of the Gribov-Zwanziger scenario~\cite{Gribov:1977wm}-\cite{Dudal:2008sp}.

In order to get a deeper understanding of the 
results arising from lattice simulations, it would be clearly
interesting to obtain information on the IR behaviour 
of the ghost and gluon propagators in as many gauges as possible,
including the background field gauge.

Indeed, the implementation of the BFM on the lattice
(for whatever value of the gauge fixing parameter) would be a long awaited leap forward~\cite{Dashen:1980vm}. 

One could in particular explore the influence
of topological non-trivial backgrounds on the
Green functions of the theory in a non-perturbative setting.
In the pioneering paper \cite{'tHooft:1976fv} it has been shown
how to compute the quantum corrections to the
classical Yang-Mills action in the presence
of a background instanton configuration in the one-loop
approximation.
The presence of zero modes
in the two-point Green function of the
quantum fields propagating in the background
prevents to carry out a straighforward 
Gaussian integration and requires a dedicated
treatment.
This technique has been extended to two-loop order
in \cite{Morris:1984zi}. 
On the other hand, a systematic procedure for 
dealing with the dependence on the background field
beyond perturbation theory is, to the best of our knowledge, still
missing.

In \cite{Binosi:2011ar} it was shown that the dependence
of the vertex functional on the background gauge field
is fixed by the Slavnov-Taylor identity. The analysis was carried
out in the background Landau gauge. 

In this paper we wish
to extend this result to an arbitrary background
$R_\xi$-gauge.
This will pave the way for the systematic
implementation of the BFM in a non-perturbative
setting in the presence of topologically
non-trivial configuration.  In this way one might be able to describe what happens when topological effects are properly taken into account, e.g. by comparing with what has been observed on the lattice when center vortices are removed from the vacuum configurations~\cite{de Forcrand:1999ms,Gattnar:2004bf}.

We will show that the dependence of the vertex functional
on the background field is fixed by a canonical
transformation w.r.t. the Batalin-Vilkovisky bracket naturally
associated with the BRST symmetry of the model \cite{Binosi:2011ar}.
The canonical transformation provides the correct way
of handling the (non-trivial) deformation of the classical
background-quantum splitting, induced by
quantum corrections.

We will then find that the dependence on the
background field can be recovered by carrying out a
suitable field redefinition, which in general
involves both the gauge and the ghost fields.

Since the method relies on symmetry requirements only,
and in particular on the ST identity in the presence
of a background field, it can be applied in any
non-perturbative computational framework which fulfills the relevant
functional identities of the model, thus easing the
matching of results obtained in different approaches
to non-perturbative QCD.

\medskip
The paper is organized as follows. In Sect.~\ref{sec.2}
we set up our notation, introduce the tree-level
vertex functional and the BV bracket generated by the BRST symmetry.
We also write the relevant functional identities of the theory
(B-equation, antighost equation, background Ward identity,
Slavnov-Taylor identity). We work in an arbitrary
background $R_\xi$-gauge.
In Sect.~\ref{sec.3} we compare the BFM in the perturbative
vs. non-perturbative frameworks.
In Sect.~\ref{sec.4} we move on to the analysis of the
constraints on the background field dependence of
the vertex functional encoded in the ST identity.
By making use of cohomological tools we will show
that the full dependence on the background connection $\hat A_\mu$
is completely fixed by the ST identity. 
This is our central result. It can be summarized in a compact
formula by means of homotopy techniques. This formula is the
basis of further applications, which we briefly outline in the Conclusions.

\section{Classical Action and Its Symmetries}\label{sec.2}

We consider Yang-Mills theory based on a semisimple gauge
group $G$ with generators $T_a$ in the adjoint representation, satisfying
\begin{eqnarray}
[ T_a, T_b] = i f_{abc} T_c \, .
\label{e.1}
\end{eqnarray}
The Yang-Mills action $S_{YM}$ is 
\begin{eqnarray}
S_{YM} = -\frac{1}{4g^2} \int d^4x \,  G_{a\mu\nu}
G_a^{\mu\nu}
\label{e.2}
\end{eqnarray}
where $g$ is the coupling constant and $G_{a\mu\nu}$ is the
Yang-Mills field strength
\begin{eqnarray}
G_{a\mu\nu} = \partial_\mu A_{a\nu} - \partial_\nu A_{a\mu}
+ f_{abc} A_{b\mu} A_{c\nu} \, .
\label{e.3}
\end{eqnarray}
We adopt a (background) $R_\xi$-gauge-fixing condition by
adding to $S_{YM}$ the gauge-fixing term
\begin{eqnarray}
S_{g.f.} & =& \int d^4x \, s \Big [ \bar c_a \Big (
\frac{\xi}{2} B_a - D_\mu[\hat A] (A-\hat A)_a \Big ) \Big ] 
\nonumber \\
& = & \int d^4x \, \Big ( \frac{\xi}{2} B_a^2
 - B_a D_\mu[\hat A] (A-\hat A)_a \nonumber \\
& & ~~~~~~~~ + \bar c_a D_\mu[\hat A] (D^\mu[A] c)_a
+ (D_\mu[A] \bar c)_a \Omega_a^\mu \Big ) \, .
\label{e.4}
\end{eqnarray}
In the above equation $\xi$ is the gauge parameter (the Landau
gauge used in \cite{Binosi:2011ar} is obtained for $\xi=0$) and $\hat A_{a\mu}$
denotes the background connection. 
$\bar c_a,c_a$ are the antighost and ghost fields respectively
and $B_a$ is the Nakanishi-Lautrup multiplier field.

We will sometimes use the notation $A_\mu = A_{a\mu} T_a$
and similarly for $\hat A_\mu, \bar c, c, B$.

The BRST differential $s$ acts on the fields of the theory as follows
\begin{eqnarray}
&& s A_{a\mu} = D_\mu[A] c_a \equiv \partial_\mu c_a + f_{abc} A_{b\mu}
 c_c \, , \nonumber \\
&& s c_a = -\frac{1}{2} f_{abc} c_b c_c \, , \nonumber \\
&& s \bar c_a = B_a \, , \quad s B_a =0 
\label{e.5}
\end{eqnarray}

$s$ is nilpotent. $\Omega_{a\mu}$ is an external source with ghost number $+1$
pairing with the background connection $\hat A_{a\mu}$ into a BRST
doublet \cite{Quadri:2002nh}
\begin{eqnarray}
s \hat A_{a\mu} = \Omega_{a\mu} \, , \qquad s \Omega_{a\mu} = 0 \, .
\label{e.6}
\end{eqnarray}
$\Omega_{a\mu}$ was introduced in \cite{Grassi:1995wr}, where it was
shown that no new  $\widehat A_\mu$- 
and $\Omega_\mu$-dependent anomalies 
can arise, as a consequence of the pairing in
eq.(\ref{e.6}).

Since the BRST transformations of the fields $A_{a\mu}$ and
$c_a$ in eq.(\ref{e.5}) are non-linear in the quantum fields,
we need a suitable set of sources,  known as antifields \cite{af,Gomis:1994he}, in order 
to control their quantum corrections.
For that purpose we finally add to the classical action the following 
antifield-dependent term
\begin{eqnarray}
S_{a.f.} = \int d^4x \, \Big ( A_{a\mu}^* D^\mu[A] c_a -
c_a^* \Big ( -\frac{1}{2} f_{abc} c_b c_c \Big ) - \bar c^*_a B_a\Big ) \, .
\label{e.6.1}
\end{eqnarray}
Although it is not necessary for renormalization purposes, we have included
in eq.(\ref{e.6.1}) the antifield $\bar c^*_a$ for $\bar c_a$.
This will allow us to treat on an equal footing all the fields of the
theory by a single Batalin-Vilkovisky (BV) bracket \cite{Gomis:1994he,Batalin:1977pb}.

We summarize in Table~\ref{tableI} the ghost charge,
statistics and dimension of the fields and antifields of the theory.

We finally end up with the  tree-level vertex functional given by
\begin{eqnarray}
\fg^{(0)} & = & S_{YM} + S_{g.f.} + S_{a.f.} \, .
\label{e.6.2}
\end{eqnarray}
\begin{table}
\begin{center}
\begin{tabular}{r||c|c|c|c|c|c|c|c|c|c|}
 & $\ A_{a\mu}\ $ &  $\ c_a\ $ & $\ \bar c_a\ $  & $\ B_a\ $ &  $\ A^{*}_{a\mu} \ $ & $\ c^{*}_a\ $  
& $\ {\bar c}_a^* \ $ & $\ B_a^* \ $ 
& $\ \widehat{A}_{a\mu}\ $ & $\ \Omega_{a\mu}\ $\\
\hline\hline
Ghost charge & 0  & 1 & -1  & 0  & -1  & -2 & 0 & -1 & 0 & 1\\
\hline
Statistics  & B & F & F  & B & F &  B & B & F & B & F\\
\hline
Dimension & 1 &  0 & 2 & 2 & 3 &  4  & 2  & 2 & 1 & 1 \\
\hline
\end{tabular} 
\caption{Ghost charge, statistics (B for Bose, F for Fermi), and mass dimension of both the $SU(N)$ Yang-Mills conventional fields and anti-fields as well as background fields and sources. \label{tableI}}
\end{center}
\end{table}

\medskip

$\fg^{(0)}$ fulfills several functional identities.
\begin{itemize}
\item the Slavnov-Taylor (ST) identity
\medskip

The ST identity encodes in functional form the invariance under the
BRST differential $s$ in eqs.(\ref{e.5}) and (\ref{e.6}). In order to set up the formalism required
for the consistent treatment of the quantum deformation of the
background-quantum splitting, it is convenient to write the
ST identity within the BV formalism.

We adopt for the BV bracket the same conventions as in~\cite{Gomis:1994he}; then, using only left derivatives, one can write
\begin{eqnarray}
(X,Y) = \int\!\diff^4x \sum_\phi
\left[ (-1)^{\epsilon_{\phi} (\epsilon_X+1)}
\frac{\delta X}{\delta \phi} \frac{\delta Y}{\delta \phi^*}
- (-1)^{\epsilon_{\phi^*} (\epsilon_X+1)}
\frac{\delta X}{\delta \phi^*} \frac{\delta Y}{\delta \phi}
\right]
\label{bracket}
\end{eqnarray}
where the sum runs over the fields $\phi = \{A_{a\mu},c_a,\bar c_a,B_a \}$ and the antifields 
$\phi^* = \{ A^*_{a\mu}, c^*_a, \bar c_a^*, B^*_a \}$. In the equations above, 
$\epsilon_\phi$, $\epsilon_{\phi^*}$ and $\epsilon_X$ represent respectively the grading of the field $\phi$, the antifield $\phi^*$ and the functional $X$.

The extended ST identity arising from the invariance
of $\fg^{(0)}$ under the BRST differential
in eq.(\ref{e.5}) and eq.(\ref{e.6}), in the presence of a background field,
can now be written as
\begin{eqnarray}
\int\!\diff^4x\, \Omega_a^\mu(x)
\frac{\delta \fg^{(0)}}{\delta \widehat A^a_\mu(x)} = 
- \frac{1}{2}\, (\fg^{(0)},\fg^{(0)}) .
\label{m.1}
\end{eqnarray}
\item the B-equation
\begin{eqnarray}
\frac{\delta \fg^{(0)}}{\delta B_a} = \xi B_a - D_\mu[\hat A] (A-\hat A)_a - \bar c^*_a\, .
\label{b.eq}
\end{eqnarray}
The B-equation  guarantees the stability of the gauge-fixing condition under radiative corrections.
Notice that the r.h.s. of the above equation is linear in the quantum fields
and thus no new external source is needed in order to define it.
It does not receive any quantum corrections.
\item the antighost equation
\begin{eqnarray}
\frac{\delta \fg^{(0)}}{\delta \bar c_a} =  D[\hat A]_\mu
\frac{\delta \fg^{(0)}}{\delta A_{a\mu}^*} - 
D_\mu[A] \Omega_{a\mu} \, .
\label{antigh.eq}
\end{eqnarray}
In the background Landau gauge one can also write an equation
for the derivative of the effective action w.r.t. the ghost
$c_a$ (also sometimes called antighost equation) \cite{hep-th/0405104}. 
This was introduced in \cite{hep-th/9804013} in the context of the 
BFM formulation of Yang-Mills theory for semi-simple gauge groups in the background 't Hooft gauge.

\item the background Ward identity
\medskip

By using the background gauge-fixing condition in eq.(\ref{e.4}),
the vertex functional $\fg^{(0)}$ becomes invariant under
a simultaneous gauge transformation of the quantum fields,
external sources and the background connection, i.e.
\begin{eqnarray}
\!\!\!\!\!\!\!\!\!\!\!\!\!\!\!\!\!\!
{\cal W}_a \fg^{(0)} & = & - \partial_\mu \frac{\delta \fg^{(0)}}{\delta 
\hat A_{a\mu}} + f_{acb} \hat A_{b\mu} \frac{\delta \fg^{(0)}}{\delta \hat A_{c\mu}} - \partial_\mu \frac{\delta \fg^{(0)}}{\delta 
\hat A_{a\mu}} + f_{acb} A_{b\mu} \frac{\delta \fg^{(0)}}{\delta A_{c\mu}} + \sum_{\Phi \in \{ B,c,\bar c \}} f_{acb} \Phi_b \frac{\delta \fg^{(0)}}{\delta \Phi_c} \nonumber \\
\!\!\!\!\!\!\!\!\!
& & 
      + f_{acb} A^*_{b\mu} \frac{\delta \fg^{(0)}}{\delta A^*_{c\mu}}
      + f_{acb} c^*_b \frac{\delta \fg^{(0)}}{\delta c^*_c} 
      + f_{acb} {\bar c}^*_b \frac{\delta \fg^{(0)}}{\delta \bar{c^*}_c}= 0 \, .
\label{w.id}
\end{eqnarray}

\end{itemize}

Several comments are in order here. First we remark
that the ST identity (\ref{m.1}) is bilinear in the vertex functional,
unlike the background Ward identity (\ref{w.id}).
Thus the relations between 1-PI amplitudes, derived by functional differentiation  of the
ST identity in eq.(\ref{m.1}),
are bilinear, in contrast with the linear ones generated 
by functional differentiation of the background Ward identity
(\ref{w.id}). 
One should notice that the background Ward identity
is no substitute to the ST identity: physical unitarity
stems from the validity of the ST identity and does not follow from the background Ward identity alone 
\cite{Ferrari:2000yp}.

Since the theory is non-anomalous, in perturbation theory 
all the
functional identities in eqs.~(\ref{m.1}), (\ref{b.eq}), (\ref{antigh.eq})
and (\ref{w.id}) are fulfilled also for the full
vertex functional $\fg$ \cite{hep-th/9807191,hep-th/9905192}. This can be proven in a
regularization-independent way by standard
methods in Algebraic Renormalization \cite{Ferrari:2000yp}.
In what follows
we assume that the same identities hold true for the 
vertex functional of the theory in the non-perturbative
regime. 

\section{Perturbative vs. Non-Perturbative Background Field Method}\label{sec.3}

The source $\Omega_{a\mu}$ was used in \cite{Ferrari:2000yp,Grassi:1995wr} in order to control
the dependence of the local cohomology $H(s|d)$ of the BRST differential on the background field $\hat A_\mu$. It guarantees that 
$H(s|d)$ in the presence of the background is isomorphic to the
cohomology $H(s|d)$ when $\hat A_\mu$ is set equal to zero \cite{Ferrari:2000yp}.
This implies that the set of local observables of the theory is not
altered by the introduction of the background field and is given
by gauge-invariant operators built out from the Yang-Mills field
strength and covariant derivates thereof \cite{Barnich:2000zw}.

In ordinary perturbation theory, i.e. when the path-integral is
carried out by expanding around the trivial vacuum
$\hat A_\mu=0$, this result is the starting point for establishing
the Background Equivalence Theorem (BET) \cite{Becchi:1999ir,Ferrari:2000yp}.
We denote by $Q_\mu$ the quantum fluctuation around the background
$\hat A_\mu$, i.e. we set
\begin{eqnarray}
A_\mu = \hat A_\mu + Q_\mu
\label{e.6.2.1}
\end{eqnarray}
According to the BET, the connected Green functions
of gauge-invariant operators can be equivalently computed
by using either of the following two connected generating functionals:
\begin{eqnarray}
&& W = \left . \fg[Q,\hat A,\Phi,\zeta]
\right |_{\hat A= 0} + \int d^4x \, J_Q Q \,  + \int d^4x \, J_\Phi \Phi \,  , \nonumber \\
&& W_{bkg} = \left . \fg[Q,\hat A,\Phi,\zeta] 
\right |_{Q=0} + \int d^4x \, J_{\hat A} \hat A \,  + \int d^4x \, J_\Phi \Phi \, .
\label{e.7}
\end{eqnarray}
In eq.(\ref{e.7})  $\fg$ denotes the 1-PI generating functional, $\Phi$ is a collective
notation for the quantum fields of the theory different than the gauge
field $Q$ and $\zeta$ is a collective notation for the external sources
coupled to local composite operators in the 1-PI generating functional.
$J_Q$ is the conjugate variable of the quantum gauge field $Q$ 
under the Legendre transform in eq.(\ref{e.7}),
$J_{\hat A}$ is the conjugate variable of $\hat A$ and 
$J_{\Phi}$ is the conjugate variable of $\Phi$.

The BET states that
\begin{eqnarray}
\left . \frac{\delta^{(n)} W}{\delta \beta_1(x_1) \dots \delta \beta_n(x_n)}
\right |_{J_Q=J_\Phi=\zeta=0} =
\left . \frac{\delta^{(n)} W_{bkg}}{\delta \beta_1(x_1) \dots \delta \beta_n(x_n)}
\right |_{J_{\hat A}=J_\Phi=\zeta=0} 
\label{e.8}
\end{eqnarray}
for any set of sources $\beta_1(x_1), \dots, \beta_n(x_n)$ coupled to gauge-invariant
operators ${\cal O}_1(x_1), \dots, {\cal O}_n(x_n)$ \cite{Ferrari:2000yp}.
Combinatorially, according to eq.(\ref{e.8})  the
connected correlator $\langle T {\cal O}_1(x_1) \dots {\cal O}_n(x_n) \rangle$ can be equivalently computed by joining  with the
$Q$-propagator 1-PI amplitudes involving gauge quantum legs, evaluated at zero background (in this case one uses the 
functional $W$), or by joining with a background propagator 1-PI amplitudes, involving background gauge legs and evaluated at zero
quantum gauge field
 (as prescribed
by the functional $W_{bkg}$).

This is a very powerful result, since it allows to evaluate physical amplitudes from 1-PI Green functions with background external legs, which are in several cases significantly simpler to compute.
 
\medskip
In a non-perturbative framework, e.g. on the lattice or 
in the approach based on the Schwinger-Dyson equations, one wishes to compute 
physical observables in the presence of a topologically
non-trivial background, i.e. the path-integral is carried
out by expanding aroung a non-trivial background
gauge field $\hat A_\mu$. 

Our aim is to provide a systematic procedure for the
determination of the dependence of the vertex functional
$\fg$ on the background field by exploiting the
functional identities of the theory only and in particular
the extended ST identity.

Since the method is based on the symmetries of the theory,
it holds independently of the particular technique
used for the non-perturbative evaluation of Green functions.
In this formalism, the relations between background and quantum
1-PI amplitudes, arising from the ST identity, yield
powerful consistency conditions that can be used as a check
of computations carried out in different non-perturbative
approaches.

\section{Canonical Transformation for the Background Dependence}\label{sec.4}

In order to control the dependence on the background connection
we start from eq.(\ref{m.1})  for the full vertex functional $\fg$:
\begin{eqnarray}
\int d^4x \, \Omega_{a\mu}(x) \frac{\delta \fg}{\delta \hat A_{a\mu}(x)} =
-\frac{1}{2} (\fg,\fg) \, .
\label{c.1}
\end{eqnarray}
By taking a derivative w.r.t. $\Omega_{a\mu}(x)$ and then setting
$\Omega_{a\mu}=0$ we get
\begin{eqnarray}
\left . \frac{\delta \fg}{\delta \hat A_{a\mu}(x)} \right |_{\Omega_\mu=0} = -
 ( \left . \frac{\delta \fg}{\delta \Omega_{a\mu}(x)} \right |_{\Omega_\mu=0} ,\left . \fg \right |_{\Omega_\mu=0}) \, .
\label{c.2}
\end{eqnarray}
This equation states that the derivative of the full vertex functional
$\fg$  w.r.t. $\hat A_{a \mu}$ at $\Omega_\mu=0$ equals the variation of 
$\fg$ w.r.t. to a canonical transformation generated by the
fermionic functional 
$ \left . \frac{\delta \fg}{\delta \Omega_{a\mu}(x)} \right |_{\Omega_\mu=0}$.

This a crucial observation. First of all
 it shows that the source
$\Omega_{a\mu}$ has a clear geometrical interpretation,
being  the source of the fermionic functional which governs
the canonical transformation \cite{Gomis:1994he} giving rise to the background field dependence.
Moreover, the dependence of the vertex functional on the
background field is designed in such a way to preserve the
validity of the ST identity (since the transformation is canonical).

In a non-perturbative setting, we can use eq.(\ref{c.2}) in order to
control the background-dependent amplitudes.
For that purpose one needs a method for solving
eq.(\ref{c.2}). An effective recursive procedure is based
on cohomological techniques.
Let us introduce the auxiliary BRST differential $\omega$
given by \cite{Binosi:2011ar}
\begin{eqnarray}
\omega \hat A_{a \mu} = \Omega_{a\mu} \, , ~~~~
\omega \Omega_{a\mu} = 0 \, ,
\label{c.5}
\end{eqnarray}
while $\omega$ does not act on the other variables of the theory.
Clearly $\omega^2=0$ and, since the pair $(\hat A_{a\mu}, \Omega_{a\mu})$ forms a BRST doublet \cite{Quadri:2002nh} under $\omega$,
the cohomology of $\omega$ in the space of local functionals
spanned by $\hat A_{a\mu}, \Omega_{a\mu}$ is trivial.

This allows us to introduce the homotopy operator $\kappa$
according to
\begin{eqnarray}
\kappa = \int d^4x \, \int_0^1 dt \, \hat A_{a\mu}(x) \lambda_t \frac{\delta} {\delta \Omega_{a\mu}(x)}
\label{c.6}
\end{eqnarray}
where the operator $\lambda_t$ acts as follows on a functional
$X(\hat A_{a\mu}, \Omega_{a\mu}; \chi)$ depending
on $\hat A_{a\mu}, \Omega_{a\mu}$ and on other variables
collectively denoted by $\chi$:
\begin{eqnarray}
\lambda_t X(\hat A_{a\mu}, \Omega_{a\mu}; \chi) = 
X(t \hat A_{a\mu}, t \Omega_{a\mu}; \chi)
\label{c.7}
\end{eqnarray}
The operator $\kappa$ obeys the relation
\begin{eqnarray}
\{ \omega, \kappa \} = {\bf 1}_{\hat A,\Omega}
\label{c.8}
\end{eqnarray}
where ${\bf 1}_{\hat A,\Omega}$ denotes the identity
in the space of functionals containing at least one $\hat A_\mu$ or $\Omega_\mu$.

Then we can rewrite the ST identity (\ref{c.1}) as
\begin{eqnarray}
\omega \fg = \Upsilon \, ,
\label{c.111}
\end{eqnarray}
where
\begin{eqnarray}
\Upsilon = - \frac{1}{2}(\fg,\fg) \, .
\label{c.112}
\end{eqnarray}
By the nilpotency of $\omega$ 
\begin{eqnarray}
\omega \Upsilon = 0 \, .
\label{c.112.1}
\end{eqnarray}
Since $\left . \Upsilon \right |_{\Omega=0}=0$,
we have from eq.(\ref{c.8})
\begin{eqnarray}
\Upsilon = \{ \omega, \kappa \} \Upsilon = \omega \kappa \Upsilon
\label{c.113}
\end{eqnarray}
Thus from eq.(\ref{c.111}) we have the identity 
\begin{eqnarray}
\omega ( \fg - \kappa \Upsilon) = 0 \, ,
\label{c.114}
\end{eqnarray}
which has the general solution
\begin{eqnarray}
\fg = \fg_0 + \omega \Xi +\kappa \Upsilon
\label{c.10}
\end{eqnarray}
with $\Xi$ an arbitrary functional with ghost number $-1$.
In the above equation $\fg_0$ denotes the vertex functional
evaluated at $\hat A_\mu = \Omega_\mu =0$ (i.e.
the set of 1-PI amplitudes with no background insertions and
no $\Omega_\mu$-legs). The second term vanishes
at $\Omega_\mu=0$ but is otherwise unconstrained. 
I.e. the extended ST identity is unable to fix
 the sector where $\Omega_\mu \neq 0$.
However this ambiguity is irrelevant if one is interested
in the 1-PI amplitudes with no $\Omega_\mu$-legs, which
are those needed for physical computations.

In practical applications 
it is convenient to expand the  term $\kappa \Upsilon$ in eq.(\ref{c.10})
according to the number
of background legs, in order to write a tower of equations
allowing to solve for the dependence on $\hat A_{a \mu}$,
recursively down to the boundary condition (i.e. the 
vertex functional at zero background) $\fg_0$.
We will discuss this point elsewhere.

In the zero background ghost sector $\Omega_\mu=0$, the $\omega \Y$ term in Eq.~(\ref{c.10}) drops out, and one is left with the result 
\begin{eqnarray}
\!\!\!\!\!\!\!\!\!\!\!
\left . \fg \right |_{\Omega=0}&=&\homo\operhs+\fg_0\nonumber \\
&=& - \left . \int\! \mathrm{d}^4x \,{\widehat{A}^a_\mu(x)}\!\int_0^1\!\!\mathrm{d}t\,\lambda_t\,\frac{\delta}{\delta_{\Omega^\mu_a(x)}}\!\int\!\mathrm{d}^4y
\left[\fg_{A^{*\nu}_b}(y)\fg_{A^{b}_\nu}(y) \right . \right . \nonumber \\
&& \qquad  \left . \left . +\fg_{c^{*b}}(y)\fg_{c^b}(y)+B^b(y)\fg_{\bar c^b}(y)\right]\right|_{\Omega_\mu=0}
\nonumber \\
&&+ \fg_0.
\label{int.rep.g}
\end{eqnarray}
In the above equation we have used the short-hand notation 
$\fg_\varphi = \frac{\delta \fg}{\delta \varphi}$.
Finally, if one is interested in the sector where ghosts are absent, the formula above further simplifies to
\begin{eqnarray}
\left . 
\fg \right |_{c=0}
&=&- \left . \int\! \mathrm{d}^4x \,{\widehat{A}^a_\mu(x)}\!\int_0^1\!\!\mathrm{d}t\,\lambda_t\!\int\!\mathrm{d}^4y
\left[\fg_{\Omega^\mu_aA^{*\nu}_b}(x,y) \fg_{A^{b}_\nu}(y)
+B^b(y)\fg_{\Omega^\mu_a\bar c^b}(x,y)\right] \right|_{\Omega_\mu,c=0}\nonumber \\
&+&\left . \fg_0 \right |_{c=0}.
\label{me-final}
\end{eqnarray}
The equation above is quite remarkable,  for it provides a representation of 
the vertex functional in the ghost-free sector that isolates
the dependence on the background gauge field~$\widehat{A}_\mu$.

Both eqs.(\ref{int.rep.g}) and (\ref{me-final}) hold in an arbitrary
background $R_\xi$-gauge. Their validity can in fact
be further extended, e.g. to nonlinear gauges, since in the
derivation of eq.(\ref{c.10}) the only condition on the
gauge-fixing functional is that it should be BRST-exact (compare
with eq.(\ref{e.4})). 

\section{Conclusions}\label{sec.5}

In the present paper we have shown that the dependence
of the vertex functional on the background field 
is completely fixed by the extended ST identity.
This result is general and only relies on the validity
of the relevant functional symmetries of the theory,
thus it can be applied in any  symmetric non-perturbative setting.
One recovers the dependence on the background gauge
field by carrying out the field redefinition associated
with the canonical transformation (\ref{c.2}). 
This entails that the classical background configuration
gets deformed by quantum corrections.

One could then try to evaluate explicitly these deformations,
e.g. for the instanton profile in SU(2) Yang-Mills theory.
Moreover, the present formalism could be applied in order to
analyze the dependence on the subtraction scale $\mu$
of  physical observables in the presence of 
a background field configuration.
It is known that such a dependence is non-trivial:
at two-loop level renormalization group-invariance of physical quantities
(like e.g. the ratio of the vacuum expectation value 
in the presence of an instanton configuration over 
the vacuum expectation value around the trivial
solution $\hat A_\mu=0$) is only achieved by a proper
treatment of the anomalous dimensions and the integration
over the collective coordinates of the instanton \cite{Morris:1984zi}.

The techniques discussed in this paper could  be used in order to derive the appropriate formulation of the Callan-Symanzik
and of the renormalization group equations in the presence
of a non-trivial background.

Another important problem which awaits to be discussed
is to see whether the present
approach can be applied in order to implement the BFM
for the well-known Cornwall-Jackiw-Tomboulis 2PI effective action~\cite{Cornwall:1974vz}.

\section*{Acknowledgments}

Useful discussions with A.~A.~Slavnov and R.~Ferrari are gratefully
acknowledged.

\end{document}